# Ion Exchange in Silicate Glasses: Physics of Ion Concentration, Residual Stress, and Refractive Index Profiles

Guglielmo Macrelli[1(*)], Arun K. Varshneya[2], and John C. Mauro[3]

[1]*Isoclima SpA-R&D Department*

*Via A.Volta 14, 35042 Este (PD), Italy*

[2]*Saxon Glass Technologies, Inc. and Alfred University*

*Alfred NY, USA*

[3]*Department of Materials Science and Engineering,*

*The Pennsylvania State University*

*University Park, Pennsylvania 16802, USA*

[*]*Corresponding Author: guglielmomacrelli@hotmail.com*

**Abstract:** A systematic review of main physical effects generated by ion exchange in silicate glasses is presented. Ion concentration distributions, residual stress profiles, and refractive index effects are discussed with particular attention on the physical and mathematical underpinnings of the ion exchange process. The study has the purpose of presenting a scientific foundation to enable future developments in the field. In this respect, the objective of this article is more educational than to present new research results. Appendixes are included to consider the detail of some specific topics without disrupting the continuity of the overall discussion. Despite the review approach of this study, some original topics are included, such as the concentration distribution with variable boundary conditions and residual stress profile with anomalies due to different relaxation mechanisms, viz., either isochoric and non-isochoric and stress driven or free energy driven. The time scale of the different relaxation mechanisms results, for some specific glass chemical compositions, in the appearance of a subsurface compression maximum progressively moving apart from glass surface and eventually turning in a reversal from compression into tensile state upon prolonged ion exchange processes. Finally, a broad discussion on optical effects induced by ion exchange is presented, paying particular attention to the possibility of experimental determination of residual stress profile.

# I. Introduction

Ion exchange ("IX") is an irreversible mass transfer process [1] in a non-equilibrium state of matter [2] such as silicate glass. The glass itself is relaxing spontaneously toward the metastable supercooled liquid state. The physical effects generated by ion exchange are highlighted by changes of the ion concentration, residual stress build-up with subsequent stress relaxation phenomena, and changes of the refractive index profile. All these effects are interconnected, and one of the purposes of this study is to present a unified description of physical effects generated by ion exchange. Such effects are used in technological applications for glass chemical strengthening [3] and for optical waveguide devices [4]. Glass chemical strengthening by ion exchange below the glass transition temperature was discovered in the early 1960s [5],[6], and waveguides devices made through the ion exchange process were subsequently developed in the 1970s [7],[8]. Strengthening technologies have been exploited in glazing for transportation (aircrafts cockpit and automotive windshields), architecture (self-bearing stairs and façades), and consumer electronics. Although ion exchange has been widely studied [4],[9],[10],[11],[12], there are still some open issues and anomalies [13],[14] in the scientific understanding. This study is limited to those ion exchange processes performed without the presence of an externally applied electric or sonic field. Other limitations of this study are the plate geometry of the glass, which is often assumed to be a semi-infinite medium. The last approximations are justified based on consideration that the diffusion length of the invading ions is typically lower than the physical dimensions of the glass article in the direction of the penetrating ions. However, this assumption may be violated for ultra-thin glass or with ultra-long ion exchange times. This study has the purpose of presenting a unified framework capturing the theoretical background of physical models for ion exchange in silicate glass, and for this reason the organization used in this study is more conceptual than historical.

## II. Kinetics – Alkali Ion Concentration

### IIa. General kinetic equations

The first effect of ion exchange is the change of concentration of alkali ions inside the glass article body. This change is due to the invasion of the glass body by alkali ions supplied by an external source (usually a molten salt bath) in a binary reaction:

$$A^+ + \overline{B^+} \rightarrow B^+ + \overline{A^+} \qquad . \tag{1}$$

Here $A^+$, $B^+$ and $\overline{A^+}, \overline{B^+}$ are the two ions in the bath and in the glass respectively; A is the invading ion and B is the host ion. The driving forces for the ion exchange process are the relative differences of the electrochemical potentials of the ions in the bath and glassy phases [15]. The equilibrium condition can be accordingly formulated as [15]:

$$\left(\mu_{\overline{A^+}} - \mu_{\overline{B^+}}\right) = \left(\mu_{A^+} - \mu_{B^+}\right);$$
$$\mu_i = \left(\frac{\partial F}{\partial c_i}\right)_{c_j, V, T} \tag{2}$$

In (2) the electrochemical potential of an ion in a material phase ($\mu_i$) is defined in terms of free energy ($F$). The modeling of the ion exchange kinetics has been extensively discussed in past scientific literature [16],[17],[18],[19],[20],[21]. The gradients of the electrochemical potentials of the ions couple ($i = A^+, B^+$) generate their fluxes [12] as follows:

$$\vec{J}_i = -C_i M_i \vec{\nabla} \mu_i \quad . \tag{3}$$

where $M_i$ is the ion mobility. Electrochemical potential can be considered in two terms: one related to pure diffusion and a second due to an external electric field $\vec{E}$. Hence, Eq. (3) can be decomposed [21],[22] introducing the diffusion coefficient $D_i$:

$$\vec{J}_i = -D_i \vec{\nabla} C_i + \frac{C_i D_i e}{fkT} \vec{E} . \tag{4}$$

Here "$e$" is the charge of a proton, $k$ is Boltzmann's constant, $T$ the absolute temperature, and $f$ is a suitable correlation factor. This last factor takes into account the difference between ion mobility when driven by successive jumps from one site to another one in the glass structure (pure diffusion) and the one driven by an external electric field. Usually an electrical neutrality condition:

$$\sum_i \vec{J}_i - \vec{J}_0 = 0 , \tag{5}$$

where $\vec{J}_0$ corresponds to the current density ($\vec{i} = e\vec{J}_0$) generated by the external electric field, and a stoichiometric conservation condition:

$$\sum_i C_i - C_0 = 0 , \tag{6}$$

where $C_0$ is the constant total concentration of mobile ions in the glass, are assumed. Applying conditions (5) and (6) to Eq. (4) allows us to write the flux equation for each ion in terms of its own gradient [21].

$$\vec{J}_i = \frac{-D_i C_0 \vec{\nabla} C_i + (1-\alpha) C_i \vec{J}_0}{C_0 - \alpha C_i} \tag{7}$$

$$\alpha = 1 - \frac{D_A}{D_B} .$$

A continuity equation connecting concentration to flux for each ion holds:

$$\frac{\partial C_i}{\partial t} + \nabla \vec{J}_i = 0 . \tag{8}$$

Expressing all quantities in terms of relative fluxes $\vec{j}_i = \vec{J}_i / C_0$ and relative concentrations $c_i = C_i / C_0$, we may finally write a diffusion equation for the ion concentration:

$$\frac{\partial c_i}{\partial t} = \frac{D\nabla^2 c_i}{1-\alpha c_i} - \frac{(1-\alpha)\vec{j}_0 \cdot \vec{\nabla} c_i - \alpha D(\vec{\nabla} c_i \cdot \vec{\nabla} c_i)}{(1-\alpha c_i)^2} \ . \tag{9}$$

In the case of absence of external electric field ($\vec{j}_0 = 0$) and for a unidimensional case, Eq. (9) for ions (A) invading the glass reduces to:

$$\frac{\partial c_A}{\partial t} = \frac{\partial}{\partial x}\left[\frac{D_A}{1-\alpha c_A}\frac{\partial c_A}{\partial x}\right]. \tag{10}$$

The coefficient $D_{AB}$, also known as the interdiffusion coefficient, can be written (see Appendix A) according to the well-known Nernst-Planck [16] expression:

$$D_{AB}(c_A, c_B) = \frac{D_A}{1-\alpha \cdot c_A} = \frac{D_A \cdot D_B}{D_A \cdot c_A + D_B \cdot c_B}. \tag{11}$$

This allows equation (10) to be finally written as

$$\frac{\partial c_A}{\partial t} = \frac{\partial}{\partial x}\left[D_{AB}\frac{\partial c_A}{\partial x}\right]. \tag{12}$$

Clearly $D_{AB}$ depends on the $x$ coordinate through its dependence on $c_A$ and $c_B$. Under some specific conditions (short ion exchange times, or when $D_A \cong D_B$, for example) it can be approximated with a constant value and the diffusion equation (12) can be approximated to the well-known and extensively studied [23],[24] diffusion equation with constant diffusion coefficient:

$$\frac{\partial c_A}{\partial t} = D_{AB}\frac{\partial^2 c_A}{\partial x^2}. \tag{13}$$

This is the diffusion equation we will consider in this study for the determination of the resulting concentration field $c(x,t)$ as a consequence of ion invasion in the glass structure. The discussion regarding the interdiffusion coefficient is well documented in the literature [16],[25],[26],[27]. The concentration dependence of $D_{AB}$ results in good agreement with predictions based on mixed-alkali model proposed by Lacharme [26].

Concentration profiles of the exchanging ions can be readily measured using instruments such as the electron microprobe, or the energy dispersive x-ray spectrometer ("EDS"), or by simple etching using dilute HF solution and performing chemical analysis of the exchanging ions in the progressively etched solution. In the electron beam instruments, one must be careful regarding mobility of charged alkali ions under the influence of the electron beam [27] and local heating effects [28]. According to Eq (11), $D_{AB}$ is concentration dependent, hence, intrinsic solutions to the diffusion equation (13) are not possible. One must resort to techniques such as Boltzmann-Matano analysis [9], to extract local diffusion coefficients from a measured concentration profile. The proposed models shall be modified to take into consideration the substantial difference between a compositionally-equivalent as-melted (CEAM) glass and the corresponding one generated by ion exchange but assuming that the interdiffusion coefficient is substantially a constant value $D$. According to the most accredited model [26], as discussed by Varshneya and Milberg [29], ion-exchanged glass in the invaded areas can be thought as a composite of stacked layers of mixed-alkali glasses with a gradually varying ratio of the alkalis. The concentration dependence of the interdiffusion coefficient can be conveniently approximated by assuming a constant value; this assumption adequately describes most ion exchange processes performed to strengthen glass articles. In the next section we will focus on some analytical solutions for the diffusion equation (13) under typical initial and boundary conditions. It should be emphasized that a differential equation provides a mathematical representation of a physical problem usually by constitutive equations (3) and (7) and imposing conservation conditions (8). This is true in the domain where the equation is valid except at the boundaries. The solution at the boundary is provided by imposing the relevant boundary condition. The definition of initial and boundary conditions guarantees the uniqueness of the solution to the physical problem.

### IIb. Solutions to the diffusion equation

The most popular concentration distribution of the invading ions in the glass structure is obtained assuming unidimensional geometry, semi-infinite medium and boundary condition of an instantaneous constant equilibrium concentration at the surface achieved at zero time of the ion exchange and so maintained for all the process duration. This last condition is effectively an ideal one, nevertheless it well represents and approximates most ion exchange processes. Unidimensional geometry and semi-infinite medium are reasonable approximations justified on the consideration of the very limited size of the penetration depth of the invading ions when compared to the physical dimension of the glass article in the diffusion direction (usually into the thickness of the glass). The initial condition is expressed as:

$$c(x,0) = f(x) \ . \tag{14}$$

We will consider two types of boundary conditions at the surface ($x=0$): a prescribed function of time, (boundary condition of first kind)

$$c(0,t) = g(t), \tag{15}$$

and a condition on the ion flux that means on the derivative of $c(x,t)$ (boundary condition of second kind):

$$\left[\frac{\partial c(x,t)}{\partial x}\right]_{x=0} = \varphi(t) \tag{16}$$

Let us consider the following initial and boundary value problem (IBVP):

$$\frac{\partial c(x,t)}{\partial t} - D\frac{\partial^2 c(x,t)}{\partial x^2} = 0$$
$$c(x,0) = 0 \tag{17}$$
$$c(0,t) = g(t)$$

This problem has classical solutions based on the Duhamel principle [23]:

$$c(x,t) = \frac{x}{2\sqrt{\pi D}}\int_0^t g(\tau)\frac{e^{-x^2/[4D(t-\tau)]}}{(t-\tau)^{3/2}}$$
$$c(x,t) = \frac{2}{\sqrt{\pi}}\int_{x/2\sqrt{Dt}}^{\infty} g(t-\frac{x^2}{4D\mu^2})e^{-\mu^2}d\mu \tag{18}$$

The equivalence between the two forms of the solution can be proved by a suitable change of variables. Taking the first line of Eq. (18) and performing a suitable change of variable:

$$\mu = \frac{x}{2\sqrt{D(t-\tau)}}; \tau = t - \frac{x^2}{4D\mu^2} \rightarrow \frac{d\mu}{d\tau} = \frac{x}{4\sqrt{D}(t-\tau)^{3/2}}$$ we arrive easily to the second form as:

$$\frac{\sqrt{\pi}}{2}g(t-\frac{x^2}{4D\mu^2})\frac{d}{d\mu}\text{erf}(\mu) = g(t-\frac{x^2}{4D\mu^2})e^{-\mu^2}$$, and we can also write the second equation (18) in the following form:

$$c(x,t) = \int_{x/2\sqrt{Dt}}^{\infty} g(t-\frac{x^2}{4D\mu^2})\frac{d}{d\mu}\text{erf}(\mu)d\mu \tag{19}$$

Applying the partial integration theorem to Eq. (19) and a back change of variables, we can find an additional form of the solution to the IBVP (17):

$$c(x,t) = g(t) - g(0)\text{erf}(\frac{x}{2\sqrt{Dt}}) - \int_0^t \dot{g}(\tau)\text{erf}(\frac{x}{2\sqrt{D(t-\tau)}})d\tau \tag{20}$$

All three forms of Eqs. (18), (19) and (20) represent the same solution, and the choice is made on the basis of computational convenience. There are two interesting examples of solutions which depend on particular choices of the $g(t)$ function. The first example is for a constant $g(t) = c_0$ as $t=0$ and kept constant for all $t>0$. This is the case of immediate concentration equilibrium achieved between glass and ion source as they are put in contact (in simple batch ion exchange process $t=0$ coincides with the time for which glass is immersed in the molten salts). The derivative of $g(t)$ under the integral of (20) is zero while $g(0)=c_0$ hence:

$$c(x,t) = c_0 - c_0 \operatorname{erf}\left(\frac{x}{2\sqrt{Dt}}\right) = c_0 \operatorname{erfc}\left(\frac{x}{2\sqrt{Dt}}\right) \tag{21}$$

This is a well-known solution of the diffusion equation. The second example is for an exponential surface condition $g(t) = 1 - e^{-t/\tau_{eq}}$. This function represents a physical situation where the equilibrium at the interface between the molten salt and the glass is not instantaneous but is progressively achieved following an exponential curve with a characteristic time $\tau_{eq}$. In this case $g(0)=0$ and $\dot{g}(t) = \dfrac{e^{-t/\tau_{eq}}}{\tau_{eq}}$ so the solution of this second example is:

$$c(x,t) = 1 - e^{-t/\tau_{eq}} - \int_0^t \frac{e^{-t/\tau_{eq}}}{\tau_{eq}} \operatorname{erf}\left(\frac{x}{2\sqrt{D(t-\tau)}}\right) d\tau \tag{22}$$

A final form of the general solution to the IBVP in Eq. (17) can be found from Eq. (20) with some algebraic manipulations:

$$c(x,t) = g(0)\operatorname{erfc}\left(\frac{x}{2\sqrt{Dt}}\right) + \int_0^t \dot{g}(\vartheta)\operatorname{erfc}\left(\frac{x}{2\sqrt{D(t-\vartheta)}}\right) d\vartheta \tag{23}$$

The analytical solutions derived represent the incoming ion concentration, $c(x,t)$, in the glass after a defined time $t$ of contact of the glass with the molten salts. A second interesting case for the solution of the diffusion equation is for mixed boundary conditions. This represents two cases, a post heat treatment after the immersion time performed at a sufficient high temperature such that ions have still a remarkable mobility, and ion exchange of the glass with an ion source of limited capacity. In this last case ions are supplied through the interface up to the capacity limit of the source, as the source is exhausted ions in the glass continue their interdiffusion because it is activated by high temperature but there is no more flux at the interface. While, for ion exchange processes for glass in a salt bath it is usually assumed a virtually infinite capacity to supply ions, we can determine a process from a coated layer deposited on the glass surface or, in general, a limited reservoir that can supply ions up to a certain time; when there are no more ions to be supplied, there are no more ions crossing the glass surface. These two ion exchange situations can be mathematically modelled by considering mixed boundary conditions. This is the case when the ion source that supplies ions to glass is limited because it is made of a deposited layer on the glass surface or because the glass has been extracted from the bath and kept at high temperatures, in these situations the conventional constant source boundary condition is no more valid. A model to handle such type of ion exchange processes can be proposed with a variable boundary condition that incorporates both a first kind constant boundary condition followed at certain time by a second kind zero flux (zero derivative) boundary condition. The above ion exchange situations can be formalized considering that up to time $\tau$ the glass article is immersed in the molten salt bath with an infinite virtual capacity of supplying ions and then the glass article is extracted and kept for a certain time $t$ at a relatively high temperature. The problem can be conveniently separated in two parts.

A) First part from time $t=0$ to time $t=\tau$

Take the diffusion Eq. (13) with an interdiffusion coefficient $D_1$ corresponding to an ion exchange temperature at temperature $T_1$:

$$\frac{\partial c_a}{\partial t} = D_1 \frac{\partial^2 c_a}{\partial x^2} \tag{24}$$

Boundary condition: $c_a(o,t) = c_s$, for $0 \leq t \leq \tau$; Initial condition: $c_a(x,0) = 0$   $x > 0$

The solution is already found as the complementary error function (21):

$$c_a(x,t) = c_s \text{erfc}\left(\frac{x}{2\sqrt{D_1 \cdot t}}\right); \quad 0 \leq t \leq \tau \tag{25}$$

B) The second part of the problem for time $\tau \leq t < \infty$

Diffusion equation is considered with an interdiffusion diffusion coefficient $D_2$ corresponding to a temperature $T_2$ different, in general, from $T_1$:

$$\frac{\partial c_a}{\partial t} = D_2 \frac{\partial^2 c_a}{\partial x^2}. \tag{26}$$

Boundary condition for no flux at the glass surface is: $\left.\frac{\partial c_a}{\partial x}\right|_{x=0} = 0$, $\tau \leq t \leq \infty$; Initial condition is the already achieved ion concentration (25) at time $\tau$: $c_a(x,\tau) = c_s \text{erfc}(\frac{x}{2\sqrt{D_1 \cdot \tau}})$.

It can be noted that the diffusion coefficient at step 1 ($D_1$) is, in general, different from the diffusion coefficient at step 2 ($D_2$). The reason for this is to keep the solution as general as possible so that step 2 can be considered, in full generality, at a different process temperature of step 1.

The interdiffusion coefficient has a strong temperature dependence $D=D(T)$ that can be conveniently represented by an Arrhenius type equation through an activation energy $H$ (J/mol):

$$D(T) = D_0 \exp\left(\frac{-H}{RT}\right), \tag{27}$$

where $R$ is the universal gas constant and $T$ is expressed in Kelvin (K).

Solution to the above problem can be found in the literature [30], [31], [32]. The proposed solutions present some issues. Reference [30] identifies the application only to post-IX thermal treatments and provides a mathematical expression of the solution not particularly amenable for numerical evaluations. Reference [31] presents a solution with a typographic error in the final expression and, finally, reference [32] treats the case only for $D_1=D_2$. A solution for different diffusion coefficients, applicable to both post ion exchange thermal treatments and ion exchange from limited sources has been presented in [33] and [34]:

$$c_a(x,t) = \frac{2c_s}{\sqrt{\pi}} \int_{x/\gamma}^{\infty} e^{-y^2} \text{erf}(ky) \, dy, \tag{28}$$

where:

$$\gamma = 2\sqrt{D_1\tau + D_2 t} \quad , \quad k = \sqrt{\frac{D_1\tau}{D_2 t}} \quad . \tag{29}$$

The overall (Step1+Step2) process time is $t'=\tau+t$. It can be demonstrated that, in the limiting case $t' \to \tau$ (that is $t \approx 0$), as expected, the solution is $c_a(x,\tau) \approx c_s \text{erfc}(\frac{x}{2\sqrt{D_1\tau}})$ that is close to the boundary condition itself. This is applicable to situations where the glass, after the process, is rapidly cooled down to a temperature where ion mobility is practically nullified. More interesting is the limiting case for large times $t \to \infty$. It can be demonstrated that, in this last limiting case, the solution is:

$$c_a(x,t) = \frac{2c_s}{\pi} \sqrt{\frac{D_1\tau}{D_2 t'}} e^{-x^2/4 D_2 \cdot t'}. \tag{30}$$

The total quantity of ions $Q$ entered into the glass during step 1 can be readily evaluated:

$$Q_{AB} = 2c_s \sqrt{\frac{D_1\tau}{\pi}} \tag{31}$$

This allows the expression of $D_1\tau$ in terms of $Q_{AB}$ and the limiting solution results:

$$c_a(x,t) = \frac{2 \cdot c_s}{\pi} \frac{\sqrt{\pi}}{2} \frac{Q_{AB}}{c_s} \sqrt{\frac{1}{D_2 t'}} e^{-x^2/4D_2 t'} = \frac{Q_{AB}}{\sqrt{\pi D_2 t'}} e^{-x^2/4D_2 t'} \tag{32}$$

It is nice to realize that this is exactly the solution for a semi-infinite medium with diffusion from an infinitesimally thin layer with zero initial concentration.

We can summarize the above discussion in the following conditions:

a) when $t' \approx \tau$ we can use the constant source solution:

$$c_a(x,t) = c_s \operatorname{erfc}\left(\frac{x}{2\sqrt{D_1 t}}\right);$$

b) when $t > 4\tau$ we can use the infinitesimally thin layer solution

$$c_a(x,t) = \frac{Q_{AB}}{\sqrt{\pi D_2 t'}} e^{-x^2/4D_2 t'} \;;$$

c) In between these times $\tau < t' < 4\tau$ we shall use the exact solution:

$$c_a(x,t) = \frac{2 c_s}{\sqrt{\pi}} \int_{x/\gamma}^{\infty} e^{-y^2} \operatorname{erf}(ky) dy \;;\; \gamma = 2\sqrt{D_1 \tau + D_2 t} \;;\; k = \sqrt{\frac{D_1 \tau}{D_2 t}}.$$

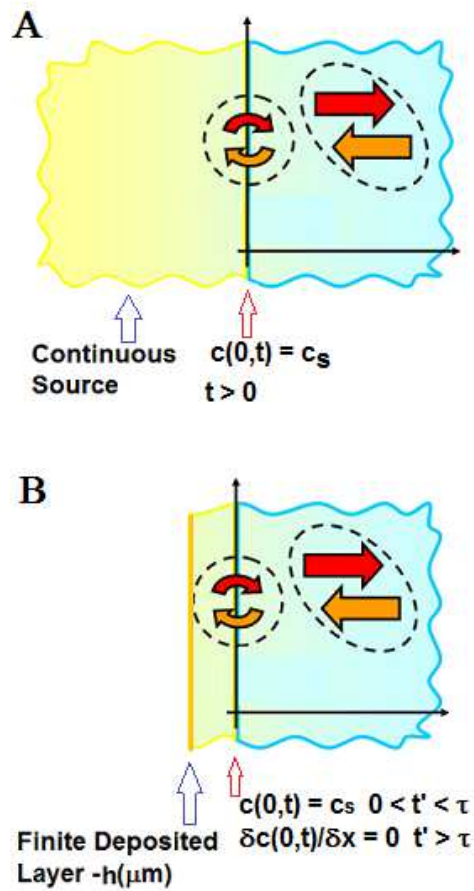

Figure 1 - Boundary Conditions: A) Continuous Source from a virtually infinite ion reservoir and B) Finite source of ions from a deposited layer

In Figure 1 the boundary condition is depicted for a continuous source with ions supplied by a virtually infinite reservoir (A) and the boundary condition (B) for a finite source of ions from a deposited thick layer. In both cases it is assumed that the interface ion source/glass reaches an almost instantaneous equilibrium condition.

### III. Mechanical effects – Residual stress

The theory outlined below is largely based on Macrelli, Varshneya and Mauro [1]. Glass strengthening by ion exchange below the glass transition temperature can be included in the class of processes based on "constrained deformation of materials" [36]. We can establish residual stress equations following the seminal idea of Cooper [37] of analogy of stress induced by concentration changes with thermoelasticity [38].

In Figure 2 a homogeneous body (A) is considered where diffusion along the coordinate *x* generates a free deformation (strain) when the body is not constrained (B). Constraints along the perpendicular coordinate limit the free strain to an average value ε (C).

This constrained deformation generates compressive stress in the near surface layers compensated by inner tensile stress.

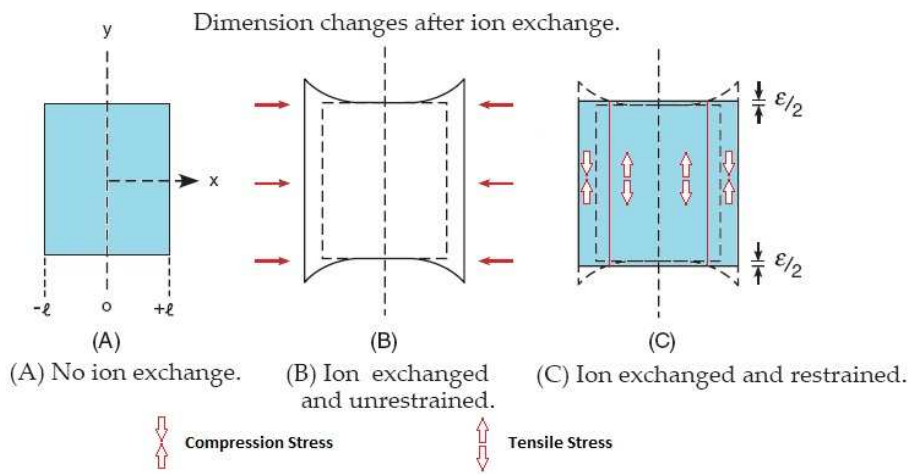

Figure 2 – Stress build-up as a consequence of constrained free expansion and by enforcement of compatibility conditions [12].

This mechanism, extended to a three dimensional configuration as depicted in Figure 3, generates an equi-biaxial stress system with a zero stress component along the direction of diffusion (making reference to figure 2, $\sigma_{xx}=0$ and $\sigma_{yy}=\sigma_{zz}$) while there are no shear stress components ($\sigma_{xy}=\sigma_{yz}=\sigma_{xz}=0$).

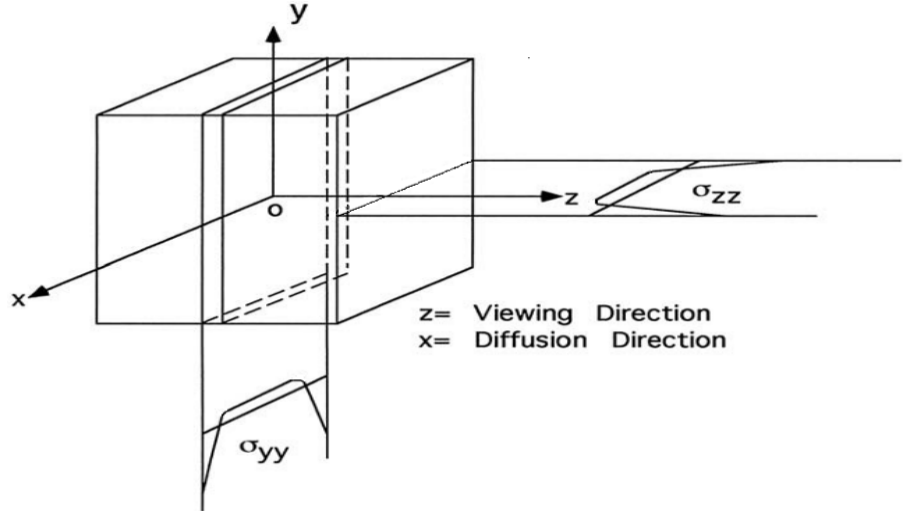

Figure 3 – Equi-biaxial stress system generated by Ion Exchange along the "*x*" coordinate.

Indicating with the indexes *(1,2,3)* the spatial coordinates (*x,y,z*) we can establish the constitutive equations for a solid with a strain induced by a concentration field as follows [39],[40]:

$$\varepsilon_{ij} = \frac{1}{E}\left[(1+\nu)\sigma_{ij} - \nu\delta_{ij}\Theta\right] + \delta_{ij}\varepsilon^c ,\qquad(33)$$

where $\delta_{ij}$ is the Kronecker delta function (=*1* if *i=j*, =0 if *i≠j*), $\varepsilon^c$ is the strain generated by the concentration of the incoming ions and:

$$\Theta = \sigma_{11} + \sigma_{22} + \sigma_{33} = 3\sigma_H ,\qquad(34)$$

where $\sigma_H$ is the hydrostatic component of the stress tensor. When *i≠j* Eq. (33) provides the shear components of the strain:

$$\varepsilon_{ij} = \sigma_{ij}\frac{1+\nu}{E} = \frac{\sigma_{ij}}{2G} ,\quad i\neq j,\qquad(35)$$

where *G* is the shear modulus. Equations (33), (34), and (35) are quite general for a solid submitted to a strain generated by concentration of incoming species. Ion exchange in glass is characterized by the following boundary condition i) and compatibility criteria ii):

i) Free strain in the x (1) direction $\varepsilon_{11} \neq 0$, which means $\sigma_{11} = 0$

ii) Enforcement of compatibility conditions which means:

Suppressed free strain in "y" (2) and "z" (3) directions (see Figure 1) so that the strain in these two directions is limited by the enforcement of compatibility criteria to a constant average strain ($\varepsilon_{ave}$). This last condition implies: $\varepsilon_{22} = \varepsilon_{33} = \varepsilon_{ave}$; which means $\sigma_{22} = \sigma_{33}$.

Indicating with $\sigma_{EB} = \sigma_{22} = \sigma_{33}$ the equi-biaxial stress generated by ion exchange (see Figure 3), according to equation (34) and by the above established boundary conditions, we can write the hydrostatic component:

$$\sigma_H = \frac{1}{3}(\sigma_{22} + \sigma_{33}) = \frac{2}{3}\sigma_{EB}, \tag{36}$$

where $\sigma_{EB}$ is the equi-biaxial stress. With above definition and applying boundary conditions for ion exchange, Eq. (33) results in the following three equations for strain:

$$\varepsilon_{11} = \frac{1}{E}[-2\nu\sigma_{EB}] + \varepsilon_C,$$

$$\varepsilon_{ave} = \frac{1}{E}(1-\nu)\sigma_{EB} + \varepsilon_C, \tag{37}$$

$$\varepsilon_{ave} = \frac{1}{E}(1-\nu)\sigma_{EB} + \varepsilon_C.$$

Eqs. (37) may be resolved for the equi-biaxial stress resulting in:

$$\sigma_{EB} = -\frac{E}{1-\nu}[\varepsilon_C - \varepsilon_{ave}]. \tag{38}$$

This is the most general relationship between the equi-biaxial stress and the strain induced by ion stuffing. The same result can be achieved by using the compatibility equations (see Appendix C).

The simplest way to express the strain induced by the invasion of incoming ions, neglecting all type of relaxation effects, is to make it linearly proportional to the incoming ions concentration $C(x,t)$:

$$\varepsilon_C(x,t) = B \cdot C(x,t). \tag{39}$$

The average strain allowed by the enforcement of compatibility criteria is :

$$\varepsilon_{ave} = \frac{1}{d}\int_0^d B \cdot C(x,t)dx, \tag{40}$$

where $d$ is the glass article thickness and $B$ is the linear network dilation coefficient also known as Cooper coefficient [13],[14] defined through the molar volume $V_m$ derivative

$$B = \frac{1}{3V_m}\frac{\partial V_m}{\partial C} \approx \frac{1}{3V_m}\frac{\Delta V_m}{C_0}. \tag{41}$$

The resulting simple stress equation, for constant $B$, is the well-known stress equation:

$$\sigma_{EB} = -\frac{E \cdot B}{1-\nu}\left[C(x,t) - \overline{C(t)}\right]. \tag{42}$$

Viscoelastic relaxation effects may be included by using the viscous relaxation function $R(t)$ [41], ($R(t)=1$ no relaxation, $R(t)=0$ fully relaxed stress).

$$[\varepsilon_C - \varepsilon_{ave}] = \int_0^t R(t-\theta)\frac{\partial}{\partial \theta}\left[B(x)C(x,\theta) - \overline{BC(\theta)}\right]d\theta. \tag{43}$$

For constant $B$ the resulting Sane-Cooper stress equation is [41]:

$$\sigma_{EB}(x,t) = -\frac{B \cdot E}{1-\nu}\int_0^t R(t-\theta)\frac{\partial}{\partial \theta}\left[C(x,\theta) - \overline{C(\theta)}\right]d\theta. \tag{44}$$

In the theory outlined above the only relaxation considered is the one due to viscoelastic components driven by isochoric shear stress. One of the most debated anomalies in the construction of a comprehensive theory of stress build-up and relaxation in ion exchange is the so called linear network dilation coefficient (LNDC) anomaly. The LNDC is just the Cooper coefficient of Eq. (41). Calculating the compressive stress generated by IX on the basis of molar volume change for CEAM compositions, we arrive at values for $\sigma_{EB}$ around 2.5 GPa against the measured values around 0.7-1 GPa. The resolution of this anomaly has been achieved by Varshneya et al. [42] and Varshneya [43] by invoking fast β-relaxation mechanisms which, in a timescale of picoseconds and nanoseconds, reduce the compressive stress to the measured one. The complexities of relaxation phenomena in IX in silicate glasses are connected to the specific chemical compositions of the involved glass families. In [1] results are discussed for specific lithium magnesium aluminosilicate glasses where compressive stress is continuously relaxed during ion exchange with sodium ions resulting in a progressive reduction of surface compression, a deeper subsurface compression maximum and, eventually in some cases, a stress reversal from compression to tension in the near surface layer. These observations, specifically the stress reversal to tensile, cannot be predicted by considering only isochoric shear stress driven relaxation. The experimental results are fully explained introducing two additional structural relaxation mechanisms:

A) A non-isochoric stress driven relaxation triggered by the almost instantaneous (picoseconds/nanoseconds time scale) compressive stress generated by the invasion of the larger ions. This mechanism is, in its nature, very similar to the β-relaxation mechanism considered in [42] for the resolution of the LNDC anomaly. In the glass family considered in [1] this mechanism contributes to an ~8% of densification of the Silica skeleton.

B) A time dependent free-energy driven structural relaxation of the silica-deficient regions which contributes to an additional ~7% of densification, resulting in the surface stress reversal from compression to tensile state.

These additional structural relaxation components are active together with the so called slow α-relaxation mechanisms (isochoric shear stress driven). In order to take into account shear stress driven viscous relaxation effects (α-relaxation), fast structural relaxation effects (β-relaxation) and free-energy driven structural relaxation effects, we introduce [1] an additional non-dimensional function $\mathcal{V}(x,t)$ in the strain term of equation (43):

$$[\varepsilon_C - \varepsilon_{ave}] = \int_0^t R(t-\theta) \frac{\partial}{\partial \theta} \left[ B(x)\mathcal{V}(x,\theta)C(x,\theta) - \overline{B\mathcal{V}C(\theta)} \right] d\theta \ . \tag{45}$$

Hence, the modified Cooper equation results:

$$\sigma_{EB}(x,t) = -\frac{E}{1-\nu} \int_0^t R(t-\theta) \frac{\partial}{\partial \theta} \left[ B(x)\mathcal{V}(x,\theta)C(x,\theta) - \overline{B\mathcal{V}C(\theta)} \right] d\theta \ . \tag{46}$$

This approach allows a formalized viscoelastic model incorporating both isochoric shear stress (deviatoric) driven components, non-isochoric hydrostatic stress components and free energy driven structural components. This is achieved by the following positions:

$$\mathcal{V}(x,t) = \mathcal{V}_0 - \Psi(t) \cdot c(x,t) \tag{47}$$

$$\Psi(x,t) = M \cdot \left[ 1 - N \cdot \exp\left(-\left(\frac{t}{\tau}\right)^b\right) \right] \tag{48}$$

Applying a mathematical transformation introduced in [34] and [44] based on the partial integration theorem, we can finally transform equation (46) into:

$$\sigma_{EB}(x,t) = -\frac{E}{1-\nu} \left[ \left[ B(x)\mathcal{V}(x,t)C(x,t) - \overline{B\mathcal{V}C(t)} \right] - \int_0^t \frac{\partial R(t-\theta)}{\partial \theta} \left[ B(x)\mathcal{V}(x,\theta)C(x,\theta) - \overline{B\mathcal{V}C(\theta)} \right] d\theta \right] \tag{49}$$

The advantage of Eq. (49) versus Eq. (46) is in the separation of the stress build-up term and the relaxation term including: α-viscous shear stress driven (isochoric) relaxation, non-isochoric hydrostatic stress driven fast β-relaxation and free energy driven structural relaxation. In Figure 4 a comparison is presented taken from [1] between uniaxial stress measured in situ during the ion exchange at the process temperature and calculated values according to the model outlined above duly corrected for the uniaxial stress according to the following equation:

$$\sigma_{UA}(x,t) = -E \cdot \left[ \left[ B(x)\mathcal{V}(x,t)C(x,t) - \overline{B\mathcal{V}C(t)} \right] - \int_0^t \frac{\partial R(t-\theta)}{\partial \theta} \left[ B(x)\mathcal{V}(x,\theta)C(x,\theta) - \overline{B\mathcal{V}C(\theta)} \right] d\theta \right]. \quad (50)$$

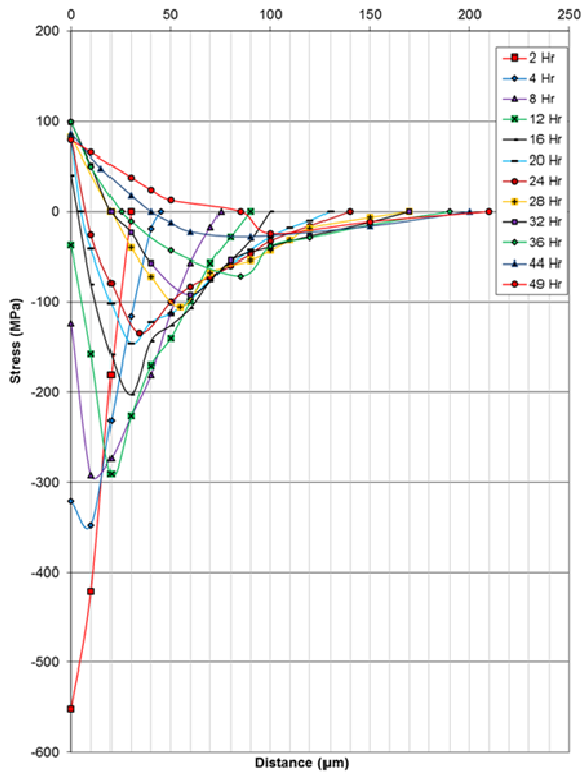
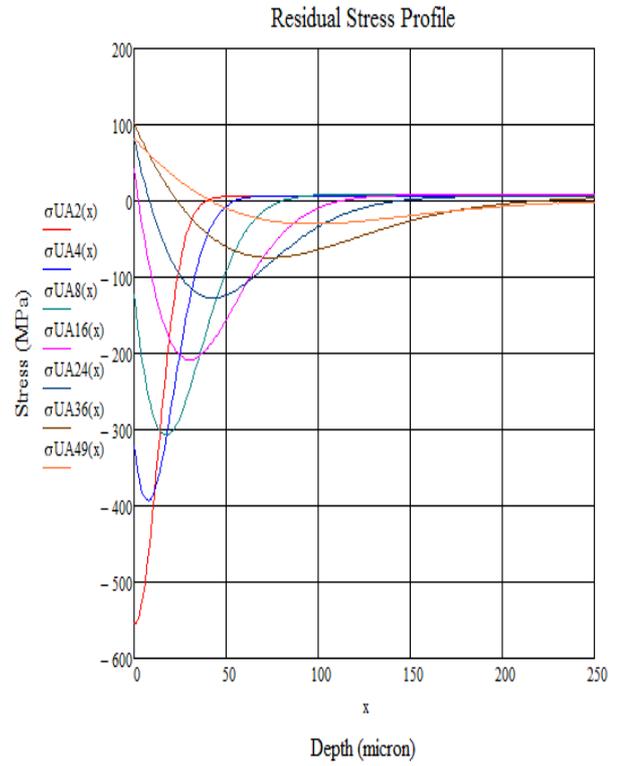

A) B)

Figure 4 – Comparison of measured A) and calculated B) stress profiles for glass 29.6Li$_2$O·9.96MgO·9.90Al$_2$O$_3$·1.15P$_2$O$_5$·49.39SiO$_2$ [1]. Values are uniaxial residual stress. Parameters for calculated values are in table 1.

It is worth noting in Figure 4 (both plots are in the same range of depth and stress) that predictions are quite in accordance with measurements. The parameters used for calculations are reported in Table 1. Recently [45] the model has been applied to a sodium-aluminosilicate glass "AS11", $13.26Na_2O \cdot 2.89K_2O \cdot 6.18MgO \cdot 0.31CaO \cdot 11.20Al_2O_3 \cdot 66.00SiO_2$. Glass samples have been immersed in a molten bath of potassium nitrate for 4 hours at a temperature of 450°C. In this case equibiaxial residual stress has been determined by differential surface refractometry (this experimental technique will be discussed in next section) using a FSM 6000 instrument manufactured by Orihara – Japan. Equibiaxial residual stress has been calculated for an ion exchange of sodium in the glass for potassium in the bath. In Figure 5 a direct comparison is presented between measured and calculated data.

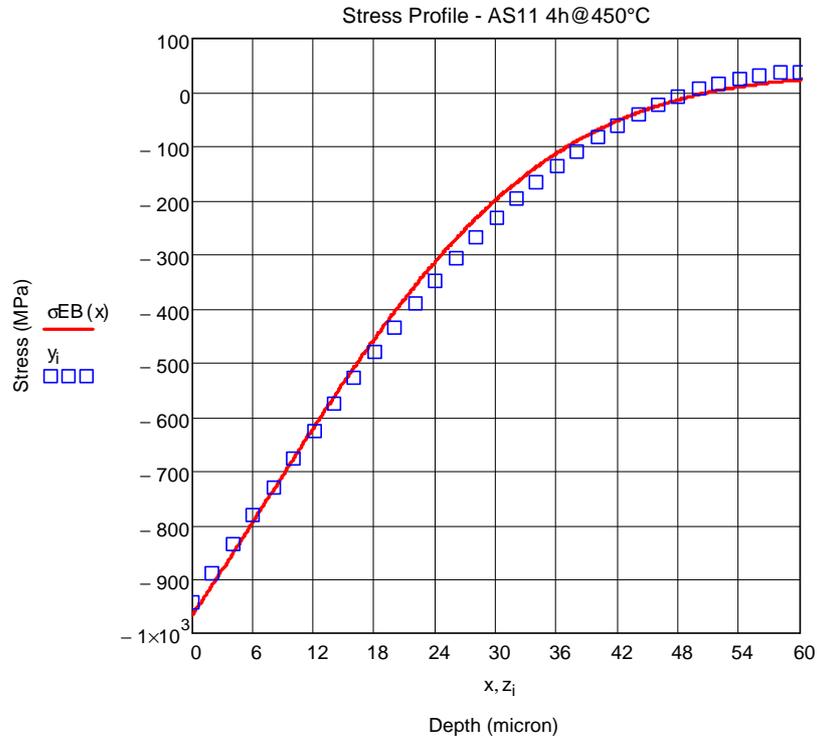

Figure 5 – Comparison of measured and calculated stress profile for $13.26Na_2O \cdot 2.89K_2O \cdot 6.18MgO \cdot 0.31CaO \cdot 11.20Al_2O_3 \cdot 66.00SiO_2$ sodium-aluminosilicate glass [45]. Values are equi-biaxial residual stress. Parameters for calculated values are in table 1.

In this last case there is practically no effect due to the free-energy structural relaxation term ($N=0$; equation (48)) and the only remarkable relaxation effects are the fast $\beta$-relaxation and the slow isochoric shear stress driven $\alpha$-relaxation effects. The above examples demonstrate the capability of the proposed mathematical model to predict residual stress values for a wide range of silicate glasses chemical compositions.

Table 1 – Parameters of calculated values according to equations (47) and (48) for the glass of Figure 4 [1] and for the glass of Figure 5 [45].

| Parameter | Values Li-Mg-Al-Si [1] - Fig.4 | Values Na-Al-Si [45] – Fig.5 |
|---|---|---|
| Glass thickness (mm) | 3 | 1.1 |
| Spatial calculation step($\mu$m) | 1 | 1 |
| Diffusion coefficient D ($\mu m^2$/s) | 0.014 – 0.030 | 0.0195 |
| Relative equilibrium surface concentration - $c(0)$ | 0.9 | 0.93 |
| Linear network dilation strain - $B \cdot C$ | 0.05984 | 0.023163 |
| Young Modulus – $E$ (MPa) | 74,800 | 71,020 |
| Poisson ratio - $\nu$ | 0.3 | 0.23 |
| Relaxation function – Stretching exponent $b$ | 3/7 | 3/7 |
| Relaxation function – Relax. Time $\tau$(s) | $0.4 \cdot 10^5 – 1.0 \cdot 10^7$ | $1.0 \cdot 10^7$ |
| $v_0$ | 0.3 | 0.63 |
| $\Psi$ coeff. $M$ | 0.50 | 0.10 |
| $\Psi$ coeff. $N$ | 0.8 | 0.0 |
| $\Psi$ coeff. $\tau_h$(s) | 32400 (9h) | N.A. |

## III. Optical effects - Refractive Index

Ion exchange generates an invasion of ions in the glass structure. In the above sections it has been shown how concentration and residual stress may be affected by this invasion. Together with the above physical effects one should consider that optical effects can be generated as a consequence of the invasion of ions with different size and polarizability [46]. The main physical property to understand how ion exchange affects optical characteristics is the refractive index. Changes in refractive index follow from three contributions: ionic polarizability, molar volume and residual stress. Let us start from the expression of the dielectric tensor $e_{ij}$ in terms of refractive index $n_{ij}$ [47]:

$$e_{ij} = \begin{pmatrix} n_{xx}^2 & n_{xy}^2 & n_{xz}^2 \\ n_{xy}^2 & n_{yy}^2 & n_{yz}^2 \\ n_{xz}^2 & n_{yz}^2 & n_{zz}^2 \end{pmatrix}. \tag{50}$$

For symmetry reasons we have only 6 independent values: $n_{xx}$, $n_{yy}$, $n_{zz}$, $n_{xy}$, $n_{xz}$, and $n_{yz}$. This allows to write the equations in a contracted form with a single index $i= (xx, yy, zz, yz, xz, xy)$. Before entering into a discussion of how ion exchange may influence its refractive index, it is worth taking a step back to understand how refractive index of glass can be modeled [49]. The electric field of light propagating through glass interacts with its polarizable species. This interaction causes the displacement of electronic charges with respect to the nuclei creating dipoles. The interaction lowers the phase velocity of light and, for this reason, the refractive index of glass is higher than the one of air. The density of the interacting species (polarizable species) is related to the glass molar volume hence it results in the dependence of refractive index on molar volume. A well-known relationship of refractive index $n$ with molar volume $V_m$ of a dielectric material (like glass) is the Lorentz and Lorenz expression [12]:

$$\frac{n^2-1}{n^2+2} = \frac{R_m}{V_m}, \tag{51}$$

where $R_m$ is the so-called molar refraction that can be expressed in terms of the polarizability $\alpha$, The Avogadro number $N_o$ and the dielectric constant of free space $\varepsilon_o$:

$$R_m = \frac{N_o}{3\varepsilon_o}\alpha. \tag{52}$$

The analysis performed in [46] allows a good expression of refractive index as a result of changes in molar refraction (polarizability effects) $\Delta R$ and molar volume $\Delta V_m$ ($n$ is the original refractive index of glass before ion exchange):

$$\Delta n_p = \frac{(n^2+2)^2}{6nV_m}\left[\Delta R - \left(\frac{n^2-1}{n^2+2}\right)\Delta V_m\right]. \tag{53}$$

In case of ion exchange of monovalent ions $A^+$ for monovalent $B^+$ ions, where $C(x)$ is the molar concentration of the invading ions, expression (51) representing only changes related to polarizability and molar volume is written as follows:

$$\Delta n_p = \frac{(n^2+2)^2}{6nV_m}\left[2(R_A - R_B)C(x) - \left(\frac{n^2-1}{n^2+2}\right)(V_{mA} - V_{mB})\right]. \tag{54}$$

The first term in (52) is the polarizability term while the second is the one due to the change of molar volume of the glass as a consequence of ion exchange. As already pointed out in the discussion about stress build up, molar volume of the exchanged glass $V_{mA}$ is not the same as the one of the glass with same composition but obtained through melt equilibrium conditions (CEAM – compositionally equivalent as-melted). This is a very important characteristic of the glass structure achieved by ion exchange essentially achieving a "forbidden state" [48].

The presence of strain coming from external actions or as a result of incompatible sources (chemical strain due to invading ion concentrations) generates a change in the dielectric tensor that can be expressed in terms of refractive index change as follows [49],[50]:

$$\Delta \begin{pmatrix} n_{xx}^{-2} \\ n_{yy}^{-2} \\ n_{zz}^{-2} \\ n_{yz}^{-2} \\ n_{xz}^{-2} \\ n_{xy}^{-2} \end{pmatrix} = \begin{pmatrix} p_{11} & p_{12} & p_{12} & 0 & 0 & 0 \\ p_{12} & p_{11} & p_{12} & 0 & 0 & 0 \\ p_{12} & p_{12} & p_{11} & 0 & 0 & 0 \\ 0 & 0 & 0 & p_{44} & 0 & 0 \\ 0 & 0 & 0 & 0 & p_{44} & 0 \\ 0 & 0 & 0 & 0 & 0 & p_{44} \end{pmatrix} \cdot \begin{pmatrix} \varepsilon_{xx} \\ \varepsilon_{yy} \\ \varepsilon_{zz} \\ \varepsilon_{yz} \\ \varepsilon_{xz} \\ \varepsilon_{xy} \end{pmatrix}. \tag{55}$$

Equation (55) is the very fundamental equation of how optical properties are influenced by mechanical strain. The equation is written in term of the strain-optical coefficients ($p_{11}$, $p_{12}$, $p_{44}$) also known as the Pockels coefficients. This equation may be transformed [50] in an explicit equation for stress (which, for this reason shall be considered derived from the fundamental one (55)):

$$\begin{pmatrix} n_{xx} \\ n_{yy} \\ n_{zz} \\ n_{yz} \\ n_{xz} \\ n_{xy} \end{pmatrix} = \begin{pmatrix} n_0 \\ n_0 \\ n_0 \\ n_0 \\ n_0 \\ n_0 \end{pmatrix} + \begin{pmatrix} K_1 & K_2 & K_2 & 0 & 0 & 0 \\ K_2 & K_1 & K_2 & 0 & 0 & 0 \\ K_2 & K_2 & K_1 & 0 & 0 & 0 \\ 0 & 0 & 0 & K_3 & 0 & 0 \\ 0 & 0 & 0 & 0 & K_3 & 0 \\ 0 & 0 & 0 & 0 & 0 & K_3 \end{pmatrix} \cdot \begin{pmatrix} \sigma_{xx} \\ \sigma_{yy} \\ \sigma_{zz} \\ \sigma_{yz} \\ \sigma_{xz} \\ \sigma_{xy} \end{pmatrix}. \qquad (56)$$

The introduced stress-optical coefficients $K_1$, $K_2$ expressed in terms of strain-optical coefficients result:

$$K_1 = \frac{n_0^3}{2E}(p_{11} - 2\nu p_{12}); K_2 = \frac{n_0^3}{2E}(p_{12} - \nu(p_{11} + p_{12})); K_3 = \frac{n_0^3}{2G} p_{44}. \qquad (57)$$

Let us put Eq. (56) in the principal axis coordinates, in this case the mixed terms in equation (56) are zeroed and it can be written:

$$\begin{pmatrix} n_{xx} \\ n_{yy} \\ n_{zz} \end{pmatrix} = \begin{pmatrix} n_0 \\ n_0 \\ n_0 \end{pmatrix} + \begin{pmatrix} K_1 & K_2 & K_2 \\ K_2 & K_1 & K_2 \\ K_2 & K_2 & K_1 \end{pmatrix} \cdot \begin{pmatrix} \sigma_{xx} \\ \sigma_{yy} \\ \sigma_{zz} \end{pmatrix}. \qquad (58)$$

Equation (58) provides the explicit equations for the refractive indices resulting from a residual stress introduced in the glass matrix :

$$\begin{aligned} n_{xx} &= n_0 + K_1\sigma_{xx} + K_2\sigma_{yy} + K_2\sigma_{zz} = n_0 + K_1\sigma_{xx} + K_2(\sigma_{yy} + \sigma_{zz}) \\ n_{yy} &= n_0 + K_2\sigma_{xx} + K_1\sigma_{yy} + K_2\sigma_{zz} = n_0 + K_1\sigma_{yy} + K_2(\sigma_{xx} + \sigma_{zz}) \\ n_{zz} &= n_0 + K_2\sigma_{xx} + K_2\sigma_{yy} + K_1\sigma_{zz} = n_0 + K_1\sigma_{zz} + K_2(\sigma_{xx} + \sigma_{yy}) \end{aligned} \qquad (59)$$

In ion exchange processes, one can identify two sources of refractive index change. The first due to polarizability and molar volume changes is described by equation (54). The second is coming from stress build-up and relaxation and is described by equation (59) that can be rewritten:

$$(\Delta n_{xx})_s = K_1\sigma_{xx} + K_2(\sigma_{yy} + \sigma_{zz})$$
$$(\Delta n_{yy})_s = K_1\sigma_{yy} + K_2(\sigma_{xx} + \sigma_{zz}) \quad . \tag{60}$$
$$(\Delta n_{zz})_s = K_1\sigma_{zz} + K_2(\sigma_{xx} + \sigma_{yy})$$

It is worth noting that refractive index change coming from polarizability and molar volume (54) is isotropic while the ones coming from stress (60) are non-isotropic. Non-isotropy of refractive index indicates birefringence effects.

Polarizability effects and stress effects are additive and can be superimposed. Hence, we can assume a total change of refractive index as follows:

$$\Delta n_x = \Delta n_p + (\Delta n_{xx})_s$$
$$\Delta n_y = \Delta n_p + (\Delta n_{yy})_s \quad . \tag{61}$$
$$\Delta n_z = \Delta n_p + (\Delta n_{zz})_s$$

The relationships providing refractive index changes as a result of stress introduced by ion exchange can be derived applying relevant boundary conditions on the stress tensor. It is interesting to evaluate equation (60) in three boundary conditions:

A) Hydrostatic stress – $\sigma_{xx} = \sigma_{yy} = \sigma_{zz} = \sigma_H$

Equations (60) result in 3 equal equations. Refractive index in all direction is the same (isotropic) and there are no birefringence (double refraction) effects:

$$(\Delta n)_s = (K_1 + 2K_2)\sigma_H \tag{62}$$

Another interesting case of boundary conditions, which is relevant for ion exchange in silicate glasses, is the equibiaxial stress:

B) Equibiaxial stress - $\sigma_{yy} = \sigma_{zz} = \sigma_{EB}$; $\sigma_{xx} = 0$ (See Figure 4) [51]:

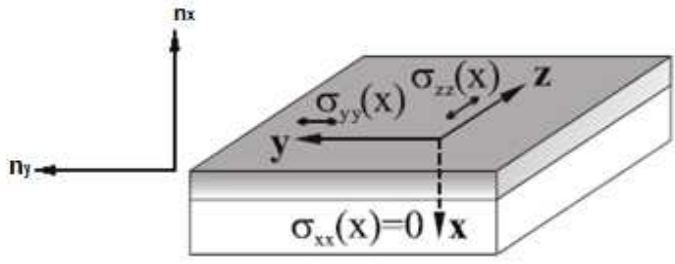

Figure 6 – Equibiaxial boundary condition representation in the principal axis coordinates.

Equation (60) results in the following index equations:

$$(\Delta n_{xx})_s = 2K_2 \sigma_{EB}$$
$$(\Delta n_{yy})_s = (K_1 + K_2)\sigma_{EB} \qquad (63)$$
$$(\Delta n_{zz})_s = (K_1 + K_2)\sigma_{EB}$$

In this case we have birefringence. If we have an electromagnetic wave travelling in the z direction, then only $n_{xx}$ (TM polarization) and $n_{yy}$ (TE polarization) are relevant (Figure 4). The interesting application of equation (63) is in the possibility of determining $\sigma_{EB}$ through measuring the difference between $\Delta n_{xx}$ and $\Delta n_{yy}$:

$$\sigma_{EB} = \frac{(\Delta n_{xx})_s - (\Delta n_{yy})_s}{K_2 - K_1} \qquad (64)$$

Third relevant case is the uniaxial stress

C) Uniaxial stress - $\sigma_{yy} = \sigma_{UA}$; $\sigma_{xx} = 0$; $\sigma_{yy} = 0$

$$(\Delta n_{xx})_s = K_2 \sigma_{UA}$$
$$(\Delta n_{yy})_s = K_1 \sigma_{UA} \qquad (65)$$
$$(\Delta n_{zz})_s = K_2 \sigma_{UA}$$

As expected, also in this case, we have birefringency and, if the wave propagates in the z-direction we can determine $\sigma_{UA}$ by the index difference:

$$\sigma_{UA} = \frac{(\Delta n_{xx})_s - (\Delta n_{yy})_s}{K_2 - K_1} \qquad (66)$$

Although Eqs. (65) and (66) appear the same, the magnitudes of $\Delta n$ are different, hence, a different result for the birefringence will be obtained between biaxial and uniaxial stresses. Stress relaxation effects can be evaluated as reduction in birefringence. In case of ion exchange resulting in an equibiaxial stress system, relaxation mechanisms are such so to merge $n_{xx}$ and $n_{yy}$ to an equal value which is $n_p$ (which will be the index value corresponding to the new composition/microstructural configuration). Performing ion exchange at very high temperature, very close to the glass transition temperature, stress build up will be relaxed in a time scale of seconds/minutes and no difference ($n_{xx}$-$n_{yy}$) will be detected in a longer time scale presenting no birefringence effects. The same will be for the hydrostatic stress condition, here we have no birefringence as the index is equal in all directions. Refractive index changes are relevant for at least two reasons: application to integrated optics [4] that is essentially the application of ion exchange to planar waveguides and the application to the determination of residual stress [51],[52],[53],[54],[55]. In both cases refractive index profiles are the main physical effects to be considered. In Figure 7, the case study for BK-7 glass (approximate wt.% 70 $SiO_2$· 3 $B_2O_3$·8.5$Na_2O$·8.5$K_2O$·3$BaO$) is shown. Concentration profile (7a) and residual stress profile (7b) have been calculated for an ion exchange process of 72 hours at 400°C in a $KNO_3$ salt bath. Calculations have been performed on the basis of the data reported in Ref. [51]. Refractive index profiles for components TE (tranverse electric), TM (transverse magnetic) and stress free (7c) have been calculated on the basis of the theory outlined above and of the photoelastic constant reported in Ref. [51]. Additionally, a focus on the merging point of TM and TE indexes is presented in Figure 7d.

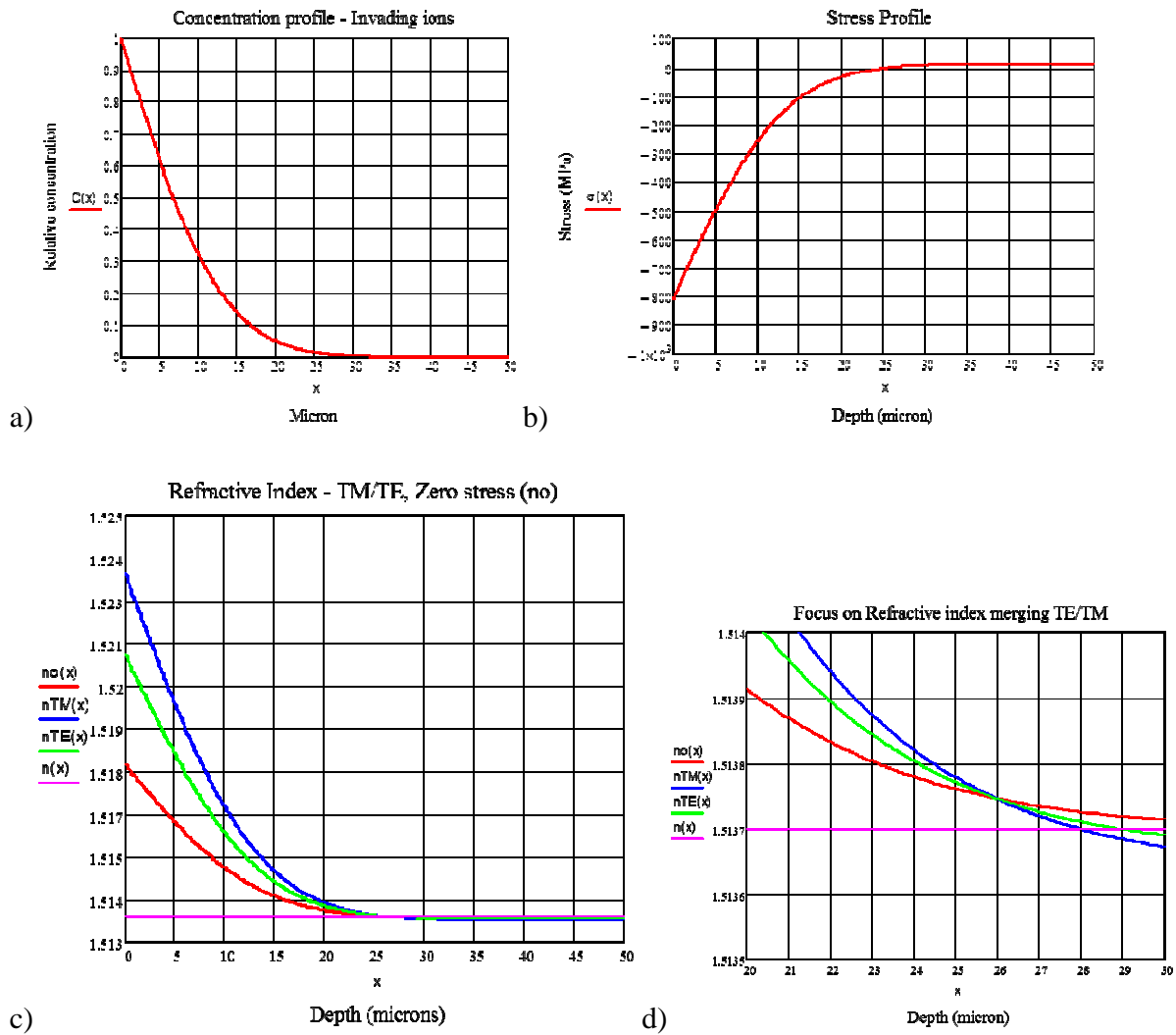

Figure 7- Concentration (7a), Stress (7b) and Refractive index profiles $n_{TM}(x)$, $n_{TE}(x)$ and $no(x)$ (7c) and focus (7d) on TM/TE merging point for BK7-glass submitted to ion exchange in KNO3 at 400°C for 72 hours [51]

A remarkable application is the determination of residual stress by using the waveguide effect generated by the index gradient and splitting the two polarization modes TE and TM [52],[53].

This can be theoretically considered from Figure 7. Indicating with $K=K_2-K_1$ the photoelastic or stress-optic coefficient, equation (64) results in:

$$\sigma_{EB} = \frac{n_{TM} - n_{TE}}{K}. \tag{67}$$

The direct measurement of the TM and TE components of the refractive index profiles allows the determination of the equibiaxial stress profile for the case where ion exchange produces refractive index profile monotonically decreasing with depth. An indirect approach is the arrangement of Figure 8. In this case a higher refractive index prism is used to allow monochromatic light to enter into a glass. The refractive index gradient generates the bending of the light modes. The effect is the generation of a double interference fringe system. The analysis of the fringes [52],[53] allows the determination of the most important and characterizing parameters of the residual stress profile that is surface compression $S_C$ (otherwise called "*CS*" or compressive stress in industry) and compression layer depth $Cd$ (called "*DOL*" or depth of layer in industry). Surface compression is defined by the value of residual stress at the glass surface: $S_C = \sigma_{EB}(x=0)$, while compression layer depth is defined by the distance from the glass surface to the inner coordinate where residual stress turns from compressive to tensile state that is where it is zero. In reference [53] an approach is presented based on the reconstruction of the refractive index profile through a mathematical procedure based on the Abel transformation. A simpler and more straightforward approach is proposed in references [52] and [55] where the interpretation of the double interference fringes system is based on the observation that ion exchange generates a gradient of the refractive index for both TM and TE modes. Bending of light beam when moving through a refractive index gradient region is a well-known physical effect [47] depicted in Figure 8.

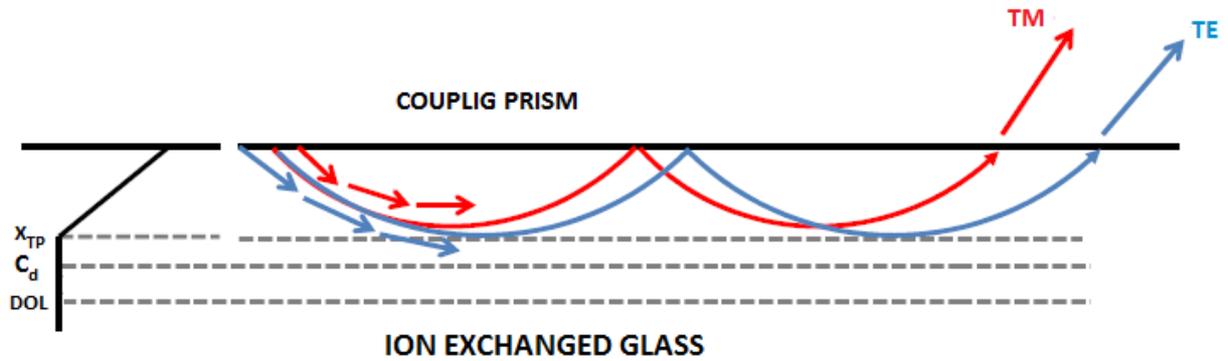

Figure 8- Bending of TM and TE modes of a light beam passing through a glass region with a graded index.

The interference condition generating the fringes can be set by the requirement that the difference between the optical path - Op - of a bent beam component (TM or TE) is a multiple of the wavelength of the monochromatic light beam ($\lambda o$):

$$O_P = \lambda_o \cdot N. \tag{68}$$

In the next discussion we will assume a linearly declining refractive index along coordinate $x$ [52]:

$$n(x) = n_s - \frac{n_s - n_o}{x_s} \cdot x \ , \ 0 \leq x \leq x_S \ , \tag{69}$$

where $n_s$ is the refractive index at the glass surface, $no$ is the refractive index of the bulk of the glass. Considering Figure 8, let us indicate with $x_{TP}$ the turning point of the light beam. It is worth noting that $x_{TP}$ it is not the depth coordinate defining the compression layer depth $Cd$ where the TM and TE modes refractive indexes merge in a single value. This is because the turning point is due to total reflection of the light beam at the critical angle and, at this point, the corresponding refractive index ($n_{TP}$) shall be higher than the refractive index without stress. The index profile TM and TE components induced by ion exchange *n(x)* merge into a single value when residual stress is null. This means that we can generally identify three points:

*$x_{TP}$* – Turning point coordinate corresponding to a refractive index *$n_{TP}$*

*Cd* – Compression layer depth coordinate corresponding to the merging of TE and TM refractive index components and to the zero stress condition.

*DOL* – Depth of ion exchanged layer corresponding to the merging of the refractive index to the bulk value before ion exchange. The above arguments requires: $DOL > Cd > x_{TP}$.

The turnaround coordinate *$x_{TP}$* is generally considered a good and acceptable approximation of the compression layer depth (*Cd*) and of *DOL* but it shall be always clear that the three values are representing different physical situations so they cannot be confused. With the above assumption of linear refractive index gradient and introducing a polar coordinates system, condition (68) allows the calculation of the optical path of the beam according to:

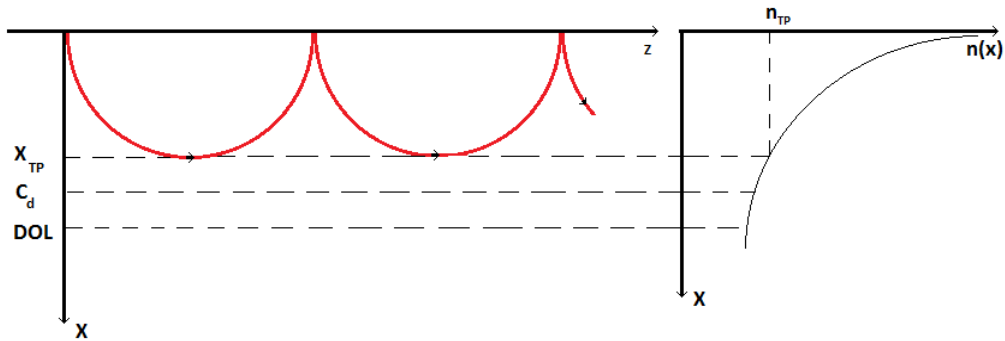

Figure 9 – The ray optic picture of light propagation in a graded index waveguide (planar) generated by ion exchange (Reworked and inspired by Figure 4.6 of reference [21]).

$$O_P = \int_C n \cdot ds = \int_0^\theta n(\gamma) \cdot R d\gamma \qquad (70)$$

Through the above expression and some additional algebraic manipulations, the compression layer depth (distance from glass surface to the turning point of the light beam) can be estimated:

$$C_d = K \frac{N}{\sqrt{(n_s - n_o)}} \quad , \tag{71}$$

where, with reference to Figure 10, $N$ is the number of the TM component fringes while the refractive index difference from surface to bulk of the glass is represented in Figure 10 by the distance between fringe A and B and $K$ is a constant:

$$K = 0.54 \cdot \frac{\lambda_o}{\sqrt{n_s}} \quad , \tag{72}$$

where $\lambda_o$ is the monochromatic light wavelength. Surface compression is easily evaluated through equation (67) by knowing the stress-optic coefficient ($K$) and the surface values of refractive indexes for TM and TE components:

$$\sigma_{EB} = \frac{n_{sTM} - n_{sTE}}{K} . \tag{73}$$

The indexes difference in (73) is represented by the distance between the corresponding surface fringes (indicated in Figure 10 as A and A') of modes TM and TE respectively. In the measuring instrument the optical system directly measures the position of the fringes. The index values to be used in equation (72) and (73) are calculated considering the sensitivity of the optical unit which depends, in turn, on the focal length of the optical system and the ratio of the refractive index of the prism of the unit to the refractive index of the glass sample.

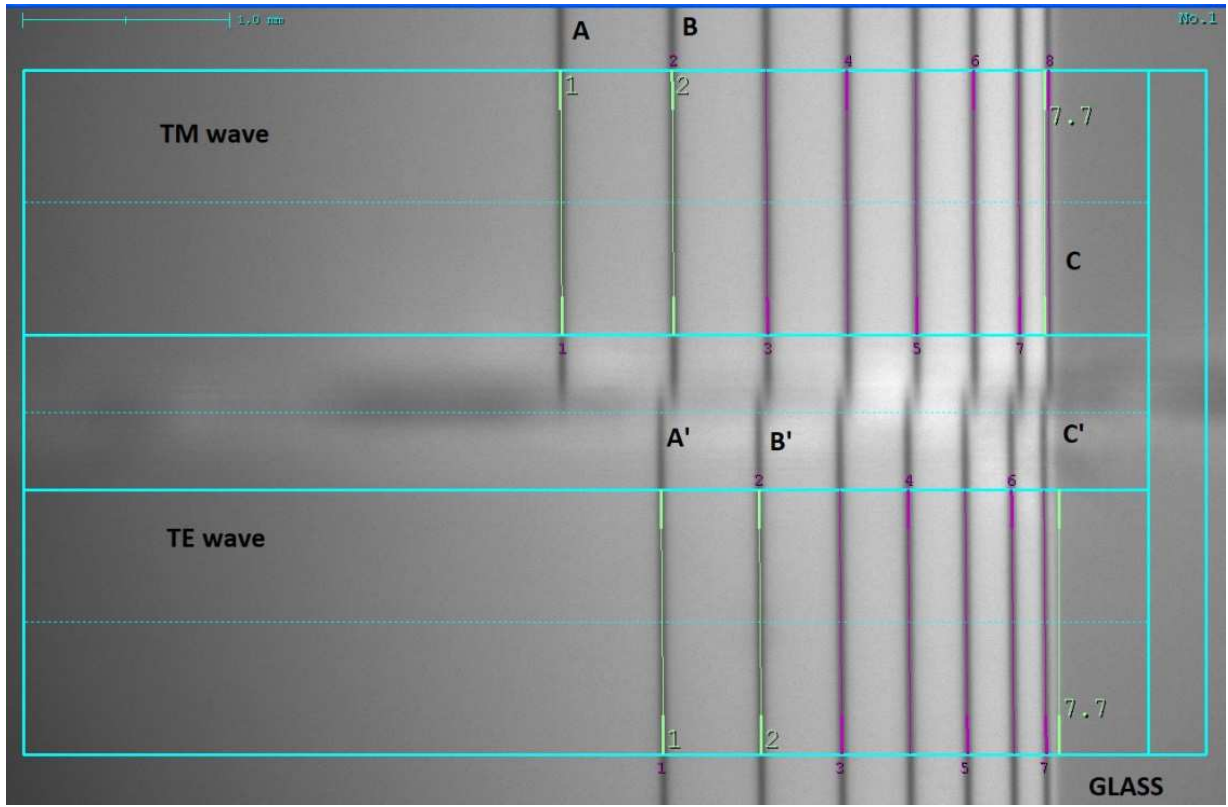

Figure 10. Fringe system image generated by a typical optically guided-wave equipment. The TM and TE components are collected and focused side by side for comparison of the fringe pattern. Fringes A and A' on the far left originate from reflection closest to the surface. Bright/Dark boundary represented by CC' is the plane in the glass where deepest internal reflection occurs (representing ion penetration depth).

The theory outlined above, that we can define differential surface refractometry, is implemented in an optical instrument commercially known as DSR or FSM [55] in experimental arrangements represented in Figure 11 (Courtesy of Roberto Dugnani).

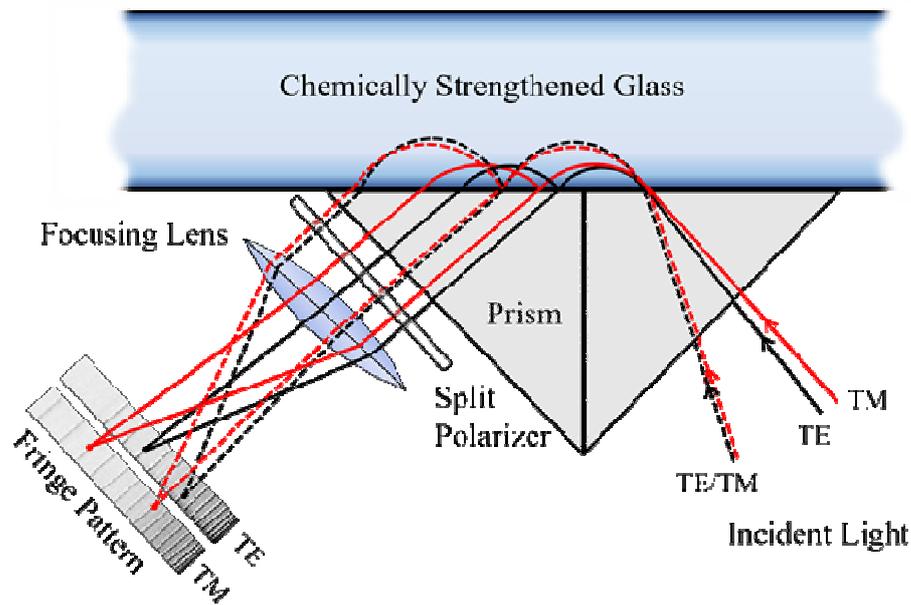

Figure 11- An instrument for the measurement of residual stress profile characteristics [52-55].

These types of instruments [56] are valuable for the determination of residual stress profile characteristics for glasses where the effect of ion exchange is to increase refractive index. Soda-lime silicates, sodium borosilicates and sodium aluminosilicate can be conveniently characterized after ion exchange by this technique. This technique does not fully work for lithium aluminosilicate [57] because of adverse polarizability effects of the lithium/sodium exchange that results in an index profile which is different to that shown in Figure 7, i.e., in a single lithium/sodium exchange the index is the lowest at the surface causing entering light rays to bend towards normal, hence, show no turnaround to exit. These lithium aluminosilicate glasses can be exposed to double or mixed ion exchange where the near surface residual stress is due to a sodium/potassium ion exchange while the inner stress profile is due to the lithium/sodium ion exchange. It has been shown [56],[57],[58],[59],[60] that residual stress profile for these glasses can be conveniently determined by DSR (for the near surface part of the stress profile) and by a Scattered Light Polarimeter ("SLP") for the deepest part of the stress profile. In the DSR instrument, light passing through a prism of higher refractive index (with a drop of intermediate refractive index oil between the prism and the glass substrate) is elevated till the condition for critical angle for refractive indices $n_{TM}$ and $n_o$ is met at the surface. A stress profile can be determined by using this apparatus by evaluating the residual stress function $\sigma_{EB}(x)$ through equation (68) identifying the corresponding fringes of TM and TE modes and evaluating the relevant refractive index differences. The possibility to measure stress by scattered light can be dated back to James Clerk Maxwell [61]. The first experimental evidence was demonstrated by Weller [62]. A significant breakthrough in this technique can be ascribed to Bateson et al. [63] who introduced laser sources in this method. The particular application we are presently discussing is based on the measurement, through scattered light intensity, of the optical retardation distribution along the path of a laser beam [57],[58]. The principle of the method is based on Rayleigh scattering generated by an obliquely incident linearly polarized light beam through a chemically strengthened glass plate. The physical mechanism can be related to refractive index fluctuations on a length scale smaller than the wavelength. These fluctuations work as dipoles scattering light in directions perpendicular to the dipole axis. The implementation of the method is made by a linearly polarized monochromatic light beam incident on a flat chemically strengthened glass plate. The phase of the incident laser light is modulated. The presence of residual stress shifts the phase of the scattered light whose intensity change is detected along its path. Indicating by R the phase shift and by x the optical path length the stress can be expressed :

$$\sigma = \frac{\lambda}{K}\frac{dR}{dx} \tag{74}$$

Where C is the stress-optic coefficient and $\lambda$ is the laser wavelength. The function R(x) is determined by the image analysis of the periodically change of the scattered light intensity at various distances from entrance point  The advantage of this method (SLP) compared with the wave guide method (DSR/FSM) is that the residual stress distribution can be determined for deep areas of the glass articles regardless of the refractive index distribution [56].

It has been found [57], [58] ,[59] that application to ion exchanged glass reduces measurement errors and resolution when an ultra narrow (50$\mu$m diameter) laser beam at a wide incidence angle (81.9°) is passed through the glass surface. The reason for having a wide incidence angle is in the significant increase in path length that can be observed. A 45° incidence angle through a 100 $\mu$m layer results in a path length of 140$\mu$m while, with a incidence angle of 81.9° the path length becomes 710$\mu$m with a factor five increase in resolution. This method is generally limited to ion exchanged glasses with compression layers larger than 20-30 mm, so it is expected that SLP method for very shallow depth of layer (DOL) is in the limit of its applicability. Details of experimental implementation and theory can be found in the cited literature [56], [57], [58],[59]. It is of particular relevance for application to lithium aluminosilicate (LAS) glasses [57],[64].

**VI. Conclusions**

Physical effects generated by ion exchange below glass transition temperature have been discussed with particular attention on the physics and mathematics needed for the development of process models. Relevant equations have been presented with their derivation from fundamental principles. Concentration distribution of the invading ions have been discussed with either constant or variable boundary conditions allowing a general expression comprising both cases as limit situations. Residual stress profiles have been presented using a formalized viscoelastic mathematical model allowing accurate prediction for most commonly used glass compositions: soda lime and sodium aluminosilicate and for specific glass compositions (lithium magnesium aluminosilicate) exhibiting a subsurface compression maximum and a reversal of compression to tensile state upon prolonged ion exchange. Refractive index effects have been discussed considering changes due to polarizability and molar volume and, ultimately, due to stress effects. The two causes of refractive index changes have been linked to fundamental physical arguments. Relevance of refractive index effects for residual stress experimental determination has been presented either by using waveguide effects of the ion exchange layer (DSR and FSM instruments) and scattered light by birefringence, which is a characteristic of an equibiaxial stress distribution.

**Acknowledgement:** One of the authors (GM) is grateful to Roberto Dugnani for enlightening discussions on the topics presented and for allowing Figure 11 to be reproduced and to Ilkay Sokmen and Nahide Ozben for allowing reproduction of Figure 5.

# APPENDIX A

Equation (10) together with the continuity equation (8) suggests that the invading ions flux can be expressed as follows:

$$J_A = -\frac{D_A}{1-\alpha \cdot c_A}\frac{\partial c_A}{\partial x} = -\frac{D_A}{1-c_A+\frac{D_A}{D_B}\cdot c_A}\frac{\partial c_A}{\partial x} =$$
$$-\frac{D_A}{\frac{D_B - D_B \cdot c_A + D_A \cdot c_A}{D_B}}\cdot\frac{\partial c_A}{\partial x} = -\frac{D_A \cdot D_B}{D_B \cdot (1-c_A) + D_A \cdot c_A}\cdot\frac{\partial c_A}{\partial x} = -\frac{D_A \cdot D_B}{D_B \cdot c_B + D_A \cdot c_A}\cdot\frac{\partial c_A}{\partial x}. \quad (A1)$$

It results the well known Nerst-Planck equation for the flux:

$$J_A = -\frac{D_A \cdot D_B}{D_B \cdot c_B + D_A \cdot c_A}\cdot\frac{\partial c_A}{\partial x}. \quad (A2)$$

The introduction of the interdiffusion coefficient $D_{AB}$ results in equation (11) and (12) that are hence justified.

# APPENDIX - B

In this appendix some calculation examples will be proposed to better understand the residual concentration effects for mixed boundary conditions.

Example A - Typical for post heating treatments after IX

1 - Initial bath immersion time = $\tau$

2 - Post Heat treatment = Let's consider two post heat treatment times: $t_1$ and $t_2$ ($t_2 \gg t_1$)

$D_1 = 8.5 \cdot 10^{-12} cm^2/s$ ; $D_2 = 2.0 \cdot 10^{-12} cm^2/s$ - $\tau = 8h$, $t_1 = 0.5h$, $t_2 = 32h$

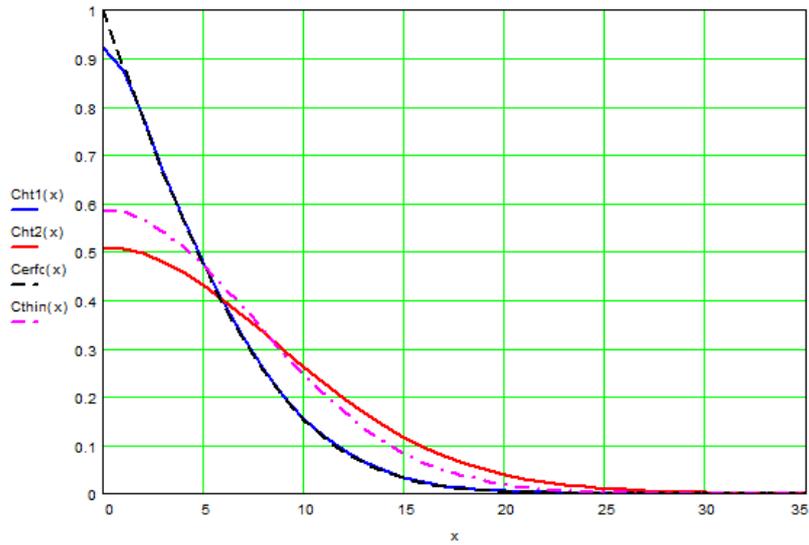

Figure B1 – Effect of post heating treatments of ion exchange processes.
Cht1 - Exact solution at time $t'=\tau+t_1$; Cht2
Exact solution at time $t'=\tau+t_2$;
Cerfc – Limit Solution for continuous source at time $\tau$,
Cthin – Limit Thin film solution at time $t'=\tau+t_2$

Example B - Typical for IX from deposited layers

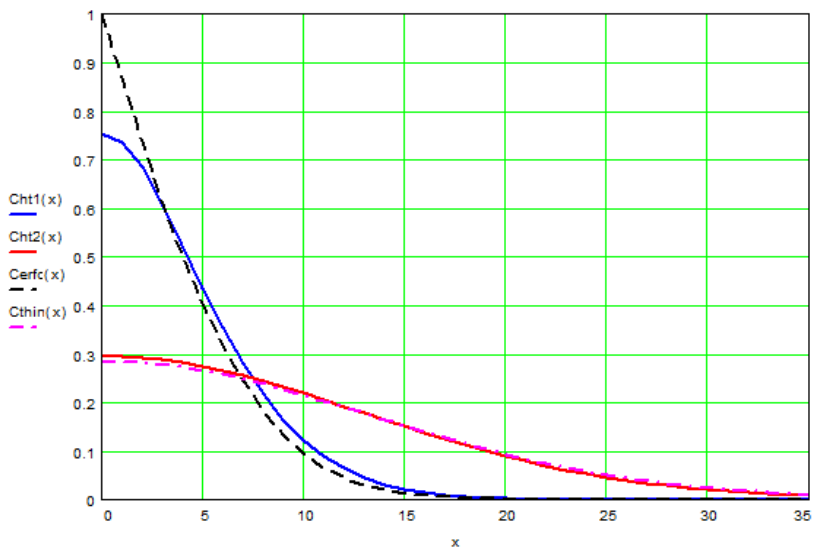

Figure B2 – Effect of ion exchange processes from a limited surface source.
Cht1 - Exact solution at time $t'=\tau+t_1$
Cht2 - Exact solution at time $t'=\tau+t_2$
Cerfc – Solution for continuous source at time $\tau$
Cthin – Thin film solution at time $t'=\tau+t$

**APPENDIX C**

The same result expressed in Eq. (38) may be achieved by using the compatibility equations [36]:

$$(1+v)\sigma_{ij,kk} + \sigma_{kk,ij} + E\left[\delta_{ij}\left(\frac{1+v}{1-v}\right)\varepsilon^C_{,kk} + \varepsilon^C_{,ij}\right] = 0. \tag{1C}$$

The tensorial form (1C) is for 6 equations that we can write in the more usual $x$, $y$ and $z$ coordinates:

$$(1+v)\nabla^2\sigma_{xx} + \frac{\partial^2\Theta}{\partial x^2} + E\left[\frac{1+v}{1-v}\nabla^2\varepsilon^c + \frac{\partial^2\varepsilon^c}{\partial x^2}\right] = 0,$$

$$(1+v)\nabla^2\sigma_{yy} + \frac{\partial^2\Theta}{\partial y^2} + E\left[\frac{1+v}{1-v}\nabla^2\varepsilon^c + \frac{\partial^2\varepsilon^c}{\partial y^2}\right] = 0,$$

$$(1+v)\nabla^2\sigma_{zz} + \frac{\partial^2\Theta}{\partial z^2} + E\left[\frac{1+v}{1-v}\nabla^2\varepsilon^c + \frac{\partial^2\varepsilon^c}{\partial z^2}\right] = 0, \tag{2C}$$

$$(1+v)\nabla^2\sigma_{xz} + \frac{\partial^2\Theta}{\partial x\partial z} + E\frac{\partial^2\varepsilon^c}{\partial x\partial z} = 0,$$

$$(1+v)\nabla^2\sigma_{yx} + \frac{\partial^2\Theta}{\partial y\partial x} + E\frac{\partial^2\varepsilon^c}{\partial y\partial x} = 0,$$

$$(1+v)\nabla^2\sigma_{yz} + \frac{\partial^2\Theta}{\partial y\partial z} + E\frac{\partial^2\varepsilon^c}{\partial y\partial z} = 0,$$

where:

$$\sigma_{kk} = (\sigma_{xx} + \sigma_{yy} + \sigma_{zz}) = \Theta \tag{3C}$$

The ion exchange boundary conditions as discussed above (reference is to Figure 3) may be expressed as follows:

$$\sigma_{yy} = \sigma_{zz} = f(x),$$

$$\sigma_{xx} = \sigma_{xz} = \sigma_{yx} = \sigma_{zy} = 0, \tag{4C}$$

$$\varepsilon^c = \varepsilon^c(x).$$

From (3C) and (4C):

$$\Theta = 2f(x). \tag{5C}$$

With the above positions, equations (2C) results in a single independent equation for $f(x)$:

$$\frac{\partial^2}{\partial x^2}\left[f(x) + \frac{E}{1-\nu}\varepsilon^c(x)\right] = 0. \tag{6C}$$

Equation (6C) has a straightforward solution:

$$f(x) = -\frac{E}{1-\nu}\varepsilon^c(x) + A_1 + A_2 x, \tag{7C}$$

where $A_1$ and $A_2$ are two constant that can be determined by imposing stress and moments equilibrium conditions:

$$\int_{-h}^{h} \sigma(x)dx = 0, \tag{8C}$$

$$\int_{-h}^{h} x\sigma(x)dx = 0. \tag{9C}$$

Imposing (8C) it results:

$$C_1 = \frac{E}{2h(1-\nu)} \int_{-h}^{h} \varepsilon^c(x)dx, \tag{10C}$$

imposing (9C) it results:

$$C_2 = \frac{3E}{2h^3(1-\nu)} \int_{-h}^{h} x\varepsilon^c(x)dx. \tag{11C}$$

Hence the residual stress is:

$$\sigma(x) = \frac{E}{1-\nu}\left[-\varepsilon^c + \frac{1}{2h}\int_{-h}^{h}\varepsilon^c dx + \frac{3x}{2h^3}\int_{-h}^{h}x\varepsilon^c dx\right]. \tag{12C}$$

If $\varepsilon^c(z)$ is an even function of $z$: $\varepsilon^c(z) = \varepsilon^c(-z)$, which holds when we have a symmetric ion exchange from both plate surfaces, equation (12C) reduces to:

$$\sigma(z) = \frac{E}{1-\nu}\left[-\varepsilon^c + \frac{1}{2h}\int_{-h}^{h}\varepsilon^c dz\right]. \tag{13C}$$

Recognizing that:

$$\varepsilon_{ave} = \frac{1}{2h}\int_{-h}^{h}\varepsilon^c dx, \tag{14C}$$

Eq. (13C) is finally:

$$\sigma(z) = -\frac{E}{1-\nu}\left[\varepsilon^c + \varepsilon_{ave}\right], \tag{15C}$$

In this way we have obtained the same result exactly as Eq. (38).

**APPENDIX D**

In this appendix a summary of mechanical (from "A" to "C") and optical ("D") conditions is reported.

A - Determination of $\sigma(x,t)$ – Equi-biaxial

Ion exchange in glass: boundary condition i) and compatibility criteria ii)
 i) Free strain in $x$ (1) direction $\varepsilon_{11} \neq 0$ which means $\sigma_{11} = 0$
 ii) Enforcement of compatibility conditions which means:
Suppressed free strain in $y$ and $z$ directions: the strain in these two directions is limited by the enforcement of compatibility criteria to a constant average strain ($\varepsilon_{ave}$). This last condition implies: $\varepsilon_{22} = \varepsilon_{33} = \varepsilon_{ave}$ which means: $\sigma_{EB} = \sigma_{22} = \sigma_{33}$.

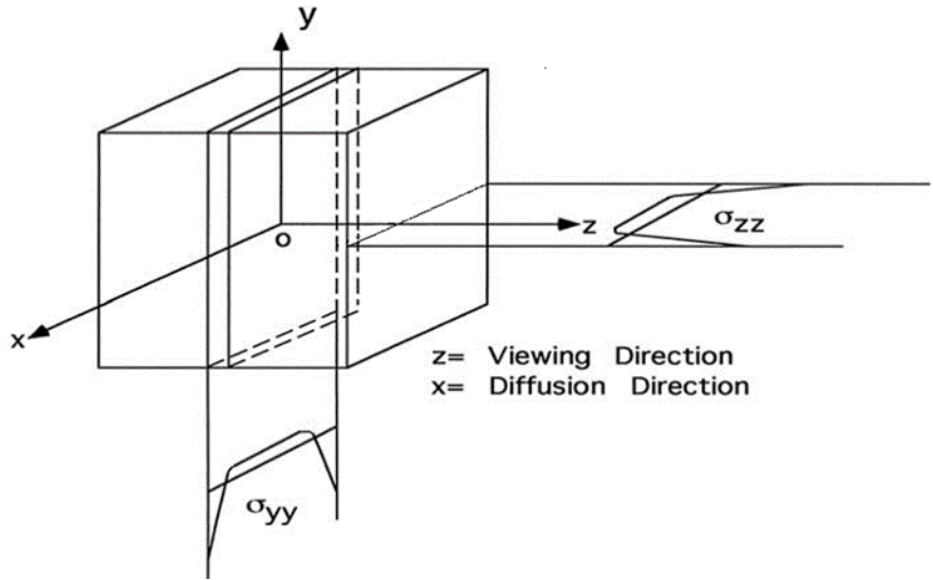

Figure 1 D – Equibiaxial stress

$$\sigma_H = \frac{1}{3}(\sigma_{22} + \sigma_{33}) = \frac{2}{3}\sigma_{EB} \tag{D1}$$

$$\sigma_{EB} = -\frac{E}{1-\nu}(\varepsilon_C - \varepsilon_{ave}) \tag{D2}$$

B - Determination of $\sigma(x,t)$ – plane stress, thin sliced samples – uniaxial stress

Ion exchange in glass: boundary condition i) and compatibility criteria ii)
  i)  Free strain in "*x*" (1) and "*z*" (3) directions $\varepsilon_{11} \neq 0$, $\varepsilon_{33} \neq 0$ that means $\sigma_{11} = \sigma_{33} = 0$.
  ii) Enforcement of compatibility conditions that means:
Suppressed free strain in "*y*" direction: the strain in this direction is limited by the enforcement of compatibility criteria to a constant average strain ($\varepsilon_{ave}$). This last condition implies: $\varepsilon_{22} = e_{ave}$ that means: $\sigma_{UA} = \sigma_{22}$

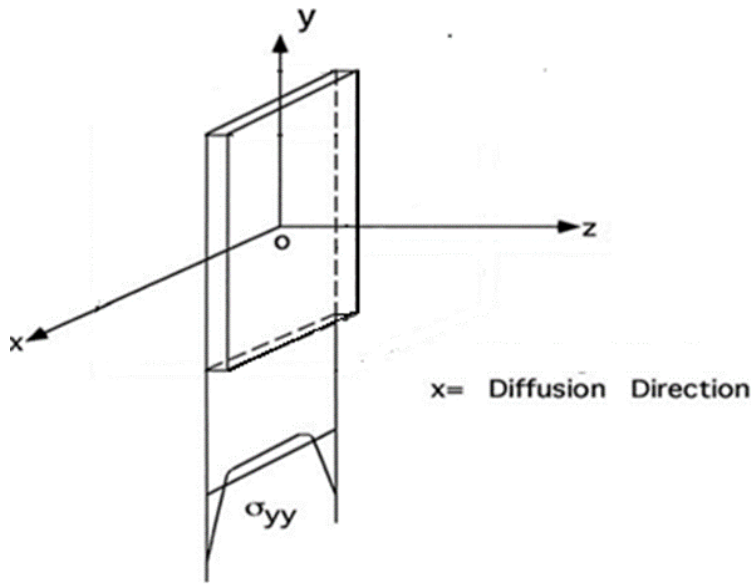

Figure 2 D – Uniaxial stress

$$\sigma_H = \frac{1}{3}\sigma_{22} \tag{D3}$$

$$\sigma_{UA} = -E(\varepsilon_C - \varepsilon_{ave}) \tag{D4}$$

C - Determination of σ(x,t) – constrained cube (NVT conditions typically used in molecular dynamics "MD" simulations) hydrostatic stress

Ion exchange in glass: boundary condition i) and compatibility criteria ii)
i) No Free strain in *x*, *y* and *z* directions
ii) Enforcement of compatibility conditions that means:

Suppressed free strain in "*x*","*y*" and "*z*" directions: the strain in these directions is limited by the enforcement of compatibility criteria to a constant average strain ($\varepsilon_{ave}$). This last condition implies: $\varepsilon_{11} = \varepsilon_{22} = \varepsilon_{33} = \varepsilon_{ave}$ that means: $\sigma_H = \sigma_{11} = \sigma_{22} = \sigma_{33}$.

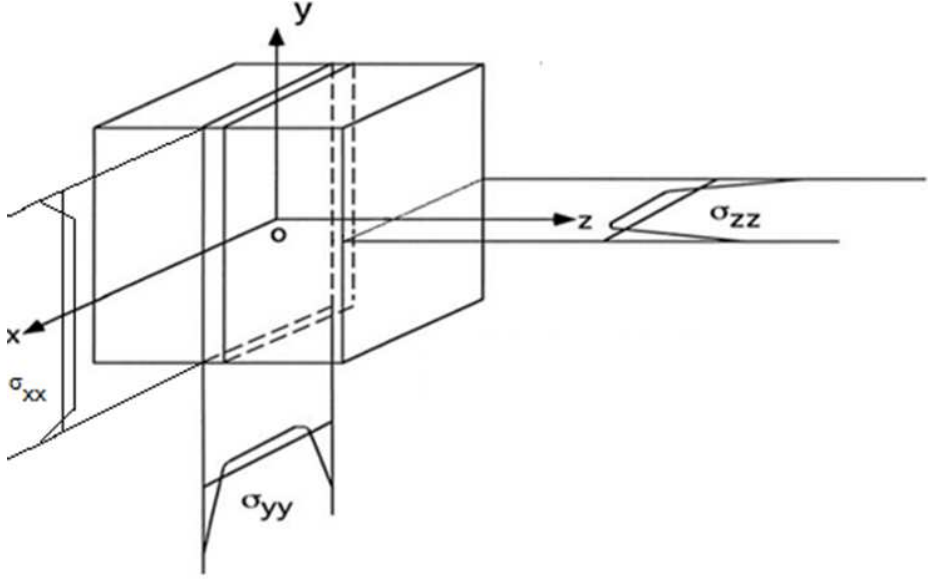

Figure 3D – Hydrostatic stress

$$\sigma_H = \sigma_{11} = \sigma_{22} = \sigma_{33} \tag{D5}$$

$$\sigma_H = -\frac{E}{1-2\nu}(\varepsilon_C - \varepsilon_{ave}) \tag{D6}$$

D - Stress and refractive index in equibiaxial conditions

$\sigma(x,t)$ and $n(x,t)$ – planar (equibiaxial)

Refractive index changes due to stress build-up:

$$n_{TM} = n_0(x) + K_1\sigma_{xx}(x) + K_2\left[\sigma_{yy}(x) + \sigma_{zz}(x)\right] \tag{D7}$$
$$n_{TE} = n_0(x) + K_1\sigma_{yy}(x) + K_2\left[\sigma_{xx}(x) + \sigma_{zz}(x)\right] \tag{D8}$$

Equibiaxial conditions:

$$\sigma_{xx}(x) = 0 \tag{D9}$$
$$\sigma_{yy}(x) = \sigma_{zz}(x) = \sigma(x) \tag{D10}$$

Expression of $K_1$ and $K_2$ in terms of Pockels Coefficients:

$$K_1 = -\frac{n_0^3}{2E}(p_{11} - 2\nu p_{12}) \tag{D11}$$

$$K_2 = -\frac{n_0^3}{2E}(\nu p_{11} + (1-\nu)p_{12}) \tag{D12}$$

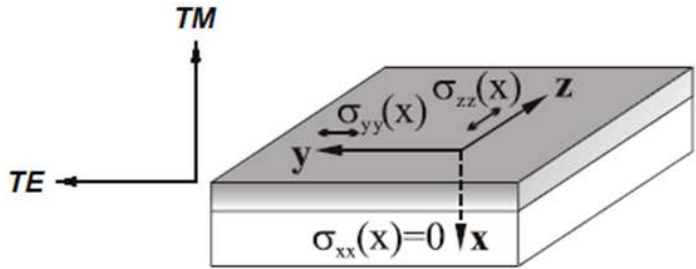

Figure 4D – Refractive index (TE and TM modes) in equibiaxial stress

$$\sigma(x) = \frac{n_{TM}(x) - n_{TE}(x)}{K_2 - K_1} \tag{D13}$$